\documentclass[a4paper,amsmath,amssymb]{revtex4}
\usepackage{xspace}
\usepackage{graphics}
\usepackage{graphicx}

\newcommand{\ie}{{\it i.e.}\xspace}
\newcommand{\eg}{{\it e.g.}\xspace}

\newcommand{\etc}{etc.\xspace}

\newcommand{\Fref}[1]{Fig.~\ref{fig:#1}}
\newcommand{\flabel}[1]{\label{fig:#1}}

\newcommand{\slabel}[1]{\label{sec:#1}}

\newcommand{\elabel}[1]{\label{eq:#1}}
\newcommand{\eref}[1]{(\ref{eq:#1})}
\newcommand{\Eref}[1]{Eq.~\eref{#1}}

\newcommand{\ave}[1]{\left\langle #1 \right\rangle\!}

\newcommand{\OC}{\mathcal{O}}

\begin{document}
\title{Equivalence of conditional and external field ensembles in
absorbing state phase transitions}
\author{Gunnar Pruessner}
\affiliation{Mathematics Institute,
University of Warwick,
Gibbet Hill Road,
Coventry CV4 7AL,
UK}
\altaffiliation[Present address: ]{Department of Mathematics,
Imperial College London, 
180 Queen's Gate, 
London SW7 2AZ}

\begin{abstract}
I comment on the relation between two sampling
methods for absorbing state models.  It is shown that a certain
ensemble without external field conditional to activity coincides with
the unconditional ensemble for sufficiently small external field.  The
actual physical processes involved are identical and the derivation of
the identity of the scaling behaviour relies on a single (established)
scaling law. While the conditional ensemble by construction does not
contain information about the system with large external field, it
contains all information about the limit of vanishing external field and
about the vicinity of the critical point: Finite size scaling as well as
critical scaling in the temperature-like variable or in (small) external
field.
\end{abstract}
\pacs{
05.70.Ln, 
05.50.+q, 
64.60.Ht, 
05.70.Jk  
}
\maketitle

\section{Introduction}
Absorbing state (AS) models suffer from the problem that the activity
necessarily ceases in any finite system in the stationarity state
\cite{Hinrichsen:2000}. In order to see a phase transition, one cannot
na{\"i}vely probe the stationary state. Two approaches are widely
accepted to overcome this problem: Either an external field is applied
that creates spontaneous activity \cite{LuebeckHeger:2003} or an
ensemble conditional to activity is considered
\cite{MarroDickman:1999} (or technically more sophisticated 
\cite{OliveiraDickman:2005}).  Both methods effectively
do away with the absorbing state altogether. The additional external
field might be physically more appealing because it resolves the problem
by introducing an external driving that appears naturally in the
corresponding field theories. On the other hand, the external field
represents a parameter that needs to be tuned in addition to the
temperature-like variable that drives the transition, because certain universal
behaviour can be obtained only for sufficiently small field
\cite{Luebeck:2004}.  The conditional ensemble has one parameter less,
but requires the selection of ``good samples'' which might appear
unphysical.

The different techniques have created a rift that goes through the
AS literature: Some authors use strong words to reject one
and support the other method. In the following it is shown, however,
that the external field ensemble coincides with a conditional
ensemble. The derivation applies to all models where the external field
triggers activity in the same way as an initial seed in the conditional
ensemble.

None of the two ensembles is superior to the other, provided they are
implemented appropriately. In fact, one can use the same implementation
and simply derive one from the other. Although the relation between the
two ensembles appears to be rather trivial, it has significant
implications in particular for the external field ensemble. Yet, so far, 
it does not seem to have entered into the AS literature.

\section{Derivation}
In the following, the relation between the moments of the activity
calculated in the different ensembles is derived. The precise definition
of the activity depends on the particular model, yet, all that matters
is that the system has an order parameter, the activity, the
instantaneous value of which vanishes for good if the system is ``left
unattended'', that is, the system hits an absorbing state after some
time. An activation mechanism can trigger a spell of activity (an avalanche) 
and these ``seeds'' are either implemented as an external field $h$ or
as the initialisation in the conditional ensemble.

Firstly, I introduce the $n$th moment of the activity in the external
field ensemble, $\ave{\rho^n}_H$. This is derived from a record of
instantaneous activity $\rho_H(t)$, see \Fref{ext_field}, in the obvious
way,
\begin{equation}
\ave{\rho^n}_H = T_0^{-1} \int_0^{T_0} dt \rho_H^n(t) \ ,
\end{equation}
where the time $T_0$ is the time span of the entire observation.

\begin{figure}
\begin{center}
\includegraphics*[width=0.8\linewidth]{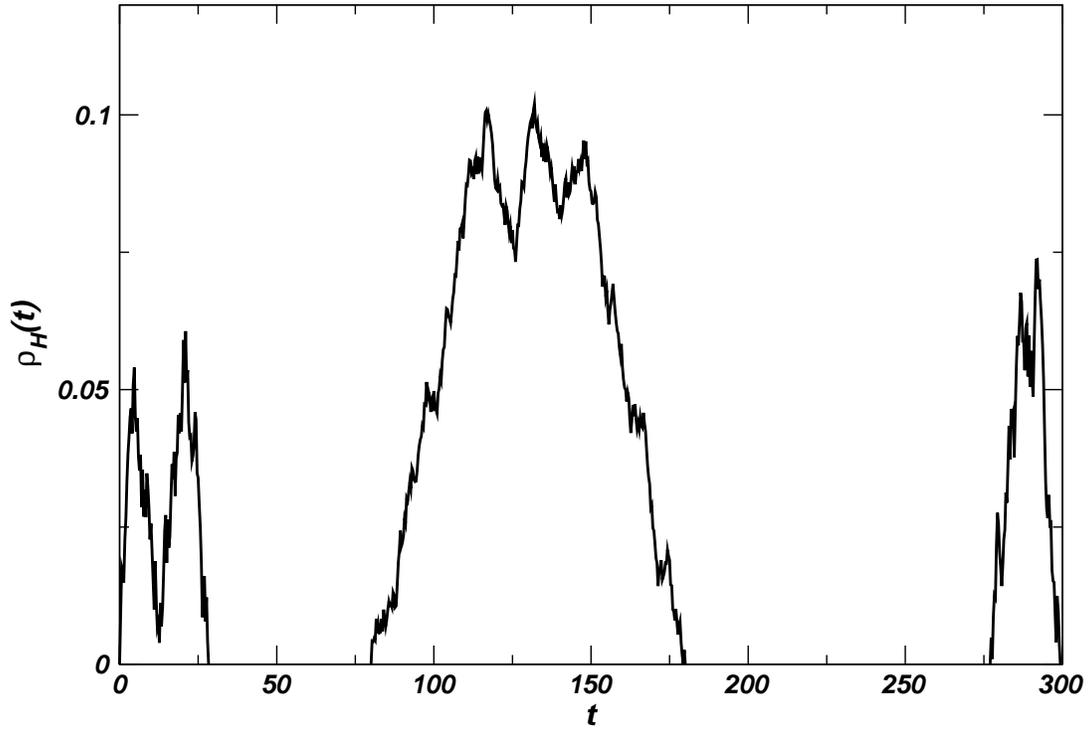}
\end{center}
\caption{\flabel{ext_field}
Cartoon of a continuous record of activity in a system with a small external field
$h$ which triggers new avalanches typically with some time of complete
quiescence between them. The total time span covered is $T_0=300$. 
The record could have been produced by
compiling the three instances shown in \Fref{conditional}.
}
\end{figure}

\begin{figure}
\begin{center}
\includegraphics*[width=0.8\linewidth]{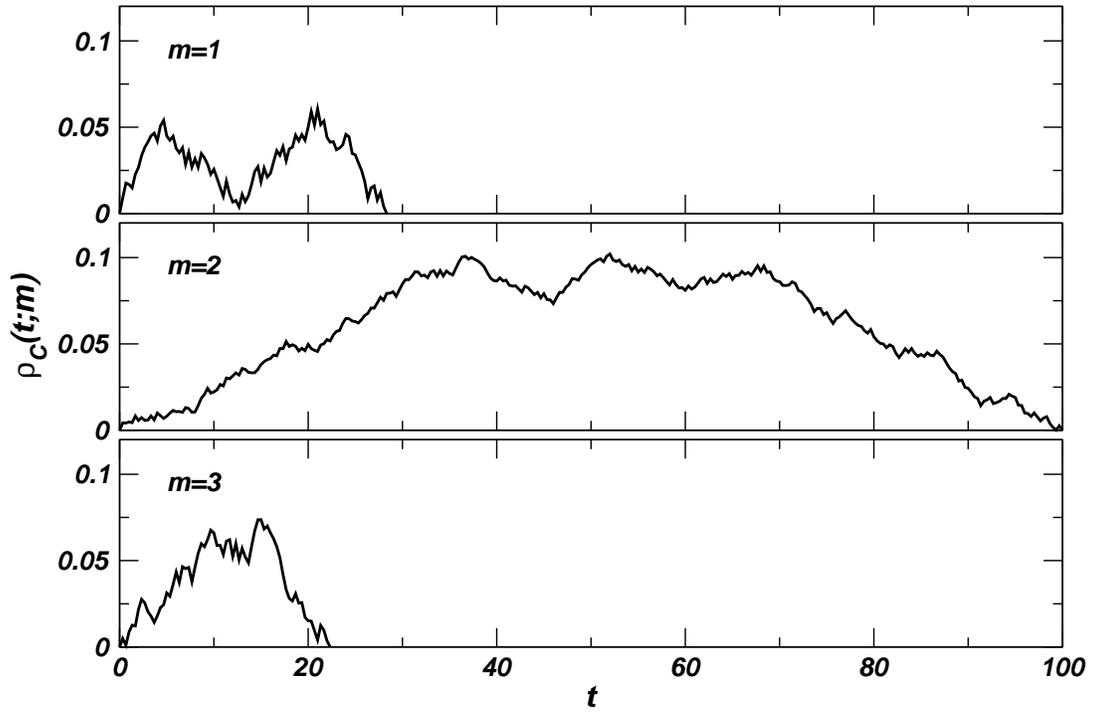}
\end{center}
\caption{\flabel{conditional}
Cartoon of three avalanches produced by placing ``seeds'' in an otherwise inactive
system. The three instances could have been taken from the continuous
record in \Fref{ext_field}.
}
\end{figure}

Secondly, I define the $n$th moment of the activity in the
conditional ensemble as the average taken conditional to the
system being active at all. \Fref{conditional} shows three instances of
conditional activity. To define it formally, $\rho_C(t;m)$ is the 
$m$th instance (\ie the $m$th run) of a spell of
activity in the system at time $t$ after initialisation at
$t=0$. If $t_m$ is the time the system is active, the moments are
\begin{equation}
\ave{\rho^n}_C =  \frac{\sum_{m=1}^M \int_0^{t_m} dt
\rho_C^n(t;m)}{\sum_{m=1}^M t_m} \ .
\end{equation}

Two technical remarks: Firstly, one should formally distinguish between
the estimate of the average from a finite sample and the ensemble
average. For brevity, this is not done here. Secondly,
$\ave{\rho^n}_C$ is often estimated slightly differently to allow for
a transient \cite{Dickman:2003}: One defines
\begin{equation}
\ave{\rho^n}_C(T) =  \frac{\sum_{m=1}^M \theta(t_m-T) \int_T^{t_m} dt
\rho_C^n(t;m)}{\sum_{m=1}^M \theta(t_m-T) (t_m-T) }
\elabel{tech_def_cond}
\end{equation}
which is a temporal average from $t=T$ to $t=\infty$ of the average
(across instances) instantaneous activity conditional to activity,
weighted by the number of systems active at time $t$. A transient can
then be established as the time $T_0$ after which the average
$\ave{\rho^n}_C(T)$ has reached a (quasi-)stationary state (see
discussion below). Transients
are commonly used in equilibrium statistical mechanics, to suppress the
effect of the initialisation. In the following, no transient is
discriminated, \ie $\ave{\rho^n}_C=\ave{\rho^n}_C(T=0)$ is used.

Both, $\ave{\rho^n}_H$ and $\ave{\rho^n}_C$ depend on the system size
$L$ and $\ave{\rho^n}_H$ is, in addition, a function of the field $h$.
\emph{However, the records $\rho_C(t)$ can be derived from $\rho_H(t)$
and vice versa, provided the field $h$ is small enough.}
This is illustrated in \Fref{ext_field} and
\Fref{conditional}: The three records in the latter Figure are derived from the
former, by restarting the clock $t$ after the $m$th spell of activity is
triggered in $\rho_H(t)$. On the other hand, given an ensemble of
avalanches $\rho_C(t;m)$, a record $\rho_H(t)$ can be constructed by
introducing gaps of vanishing activity between spells of activity taken
from $\rho_C(t;m)$. Without knowing whether the $\rho_C(t;m)$ are
independent for different $m$, they have to be compiled sequentially.
This property, however, will not be used in the following.

To construct the basic observables $\rho_H(t)$ from $\rho_C(t;m)$
correctly, one would also need to know the distribution of waiting
times. This can, in principle, depend on the configuration of the system
and might therefore vary if the system has more than one absorbing
state. However, in the following only the average waiting time will
enter.

If the two observables are derived from each other in this way, one
has by construction
\begin{equation}
\ave{\rho^n}_H =
T_0^{-1} \int_0^{T_0} dt \rho_H^n(t) =
T_0^{-1} \sum_{m=1}^M \int_0^{t_m} dt \rho_C^n(t;m) =
\frac{\sum_{m=1}^M t_m}{T_0} \ave{\rho^n}_C  \ .
\end{equation}
For a large number of avalanches $M$ the pre-factor $T_0^{-1}
\sum_{m=1}^M t_m$ converges to $\ave{w_m}_H^{-1} \ave{t_m}_C$ where
$\ave{w_m}_H$ is the average time that passes between two avalanches
being triggered in the external field ensemble and $\ave{t_m}_C$ is the
average duration of an avalanche. 

In order to derive $\ave{w_m}_H$ given $h$ one needs to know how the
external field is implemented. Provided it operates as a Poisson process
throughout the system, one has $\ave{w_m}_H=(Vh)^{-1}$ where $V$ is the
volume of the system and $h$ is the field (density) or flux. It is also
clear that the mechanism to trigger avalanches used in the conditional
ensemble must correspond to the external field operating in the system
with field. This scenario is the case for the vast majority of models
considered in absorbing state phase transitions. This is particularly
obvious if only one absorbing state is present, so that single
``seeds'', which arrive independently, uniformly and with constant rate
everywhere in the system,  trigger avalanches. 

Accepting the Poissonian nature of the external field, one arrives at
the central result
\begin{equation}
\ave{\rho^n}_H(\delta p,h;L) =
h V(L) \ave{t_m}_C(\delta p;L) \ave{\rho^n}_C(\delta p;L)  \ ,
\elabel{triviality}
\end{equation}
valid asymptotically for sufficiently small field. In \Eref{triviality}
all dependences of the individual observables on the parameters
have been stated explicitly. 
The
linearity in small field of all momenta 
$\ave{\rho^n}_H(\delta p,h;L)\propto h$ is consistent with the literature, \eg
\cite{Luebeck:2004} but see \footnote{It remains unclear why one of 17
models shown in Figures 33 and 47 of \cite{Luebeck:2004} fails to
display the linearity.}. 
The parameter $\delta p$ is the
temperature-like variable that drives a transition at $\delta p=0$,
subject of the following considerations.

\section{Discussion}
\slabel{discussion}
The above construction requires the field $h$ to be so small that no new
seeds arrive while an avalanche is running, or, more specifically, that
the statistical weight of these events is negligible. Otherwise
$\ave{\rho^n}_C$ derived from $\ave{\rho^n}_H$ would depend on the
external field, which has been excluded in \Eref{triviality} and is
crucial for the following derivations. In other words, the derivation
relies on a separation of time scales; in this sense $\ave{\rho^n}_C$ is
an observable as would be used in Self-Organised Criticality
\cite{Jensen:1998}, while $\ave{\rho^n}_H$ represents the
``instantaneous activity''
\cite{VespignaniZapperi:1997,PetersPruessner:2006}

Therefore, \Eref{triviality} can only be expected to hold if $(Vh)^{-1}$
is large compared to $\ave{t_m}_C$. This is the case for sufficiently
small field at given system size, \ie in particular for finite size
scaling where the system size is fixed before the limit $h\to0$ is taken.
The condition is also met in the inactive phase, $\delta_p<0$, where the
average duration of the avalanches converges with increasing system
size, so that, again, the average duration can be chosen to be small
compared to the time scale set by the external field. In this case, the
thermodynamic limit can be taken before $h\to0$. However, in the active
phase, the average duration diverges extremely fast with system size, so
that the external field can never be sufficiently small when the
thermodynamic limit is taken first. 

Because the external field $h$ appears only as a pre-factor on the RHS
of \Eref{triviality}, it is very simple to derive the scaling of
$\ave{\rho^n}_H(\delta p,h;L)$ from the scaling of $\ave{t_m}_C(\delta
p;L)$ and $\ave{\rho^n}_C(\delta p;L)$ assuming, naturally, $V(L)=a_V
L^d$, where $d$ is the spatial dimension of the system. 
The converse, \ie the derivation of the scaling of
$\ave{\rho^n}_C(\delta p;L)$ from the scaling of $\ave{\rho^n}_H(\delta
p,h;L)$, is not as simple and therefore shown explicitly in the
following. 

Standard scaling assumptions suggest \cite{Luebeck:2004}
\begin{equation}
\ave{\rho^n}_H(\delta p,h;L) = a_n \lambda^{-\beta_n} R_n^H(a_p \delta p
\lambda, a_h h \lambda^\sigma, a_L L \lambda^{-\nu_\perp} )
\elabel{scaling_H}
\end{equation}
with non-universal metric factors \cite{PrivmanHohenbergAharony:1991} $a_n, a_p$ \etc,  
arbitrary (scaling) parameter $\lambda$, scaling function $R_n^H$ and
the usual critical exponents $\sigma$ (field exponent), $\nu_\perp$
(spatial correlation length exponent) and $\beta_n$, the latter being
$\beta_1=\beta$ (order parameter exponent) for 
the first moment and $\beta_n=n\beta$ if gap-scaling
applies \cite{PfeutyToulouse:1977}.
Comparing \Eref{scaling_H} to \Eref{triviality} implies that $R_n^H$ is asymptotically
linear in the (small) external field
\begin{equation}
R_n^H(a_p \delta p,a_h h,a_L L) = a_h h
\tilde{R}_n(a_p \delta p , a_L L) \ ,
\elabel{linear_Rn}
\end{equation}
where $\tilde{R}_n$ must obey \Eref{scaling_H} as well, \ie 
$\tilde{R}_n(a_p \delta p, a_L L) = \lambda^{\sigma-\beta_n}
\tilde{R}_n(a_p \delta p \lambda, a_L L \lambda^{-\nu_\perp})$.

\Eref{linear_Rn} implies that the susceptibility $\chi_H(\delta
p,h;L)=\frac{\partial \ave{\rho}_H}{\partial h}$, or the derivative of
any moment $\ave{\rho^n}_H$ with respect to $h$,  is independent from
the external field under the conditions stated earlier.

To derive the scaling of $\ave{\rho^n}_C(\delta p;L)$, one needs the
scaling of $\ave{t_m}_C(\delta p, L)$ which is well understood
\cite{Luebeck:2004},
\begin{equation}
\ave{t_m}_C(\delta p, L) = a_t T(a_p \delta p, a_L L)
= a_t \lambda^{\nu_\parallel(1-\delta) } 
T(a_p \delta p \lambda, a_L L \lambda^{-\nu_\perp} )
\elabel{t_scaling}
\end{equation}
where $\nu_\parallel$ is the temporal correlation length exponent,
$\delta$ is the survival exponent and 
$\nu_\parallel (1-\delta)=\sigma - d \nu_\perp$, 
which will be crucial for the next step.  Combining
Eqns.~\eref{triviality}, \eref{scaling_H}, \eref{linear_Rn} and
\eref{t_scaling} produces the scaling the scaling of the conditional
moments:
\begin{equation}
\ave{\rho^n}_C(\delta p;L)=\frac{a_h a_n a_L^d}{a_V a_t } \lambda^{-\beta_n}
R_n^C(a_p \delta p \lambda, a_L L \lambda^{-\nu_\perp}) \ ,
\elabel{scaling_C}
\end{equation}
where $R_n^C(x, y)=\tilde{R}_n(x,y)/(y^d T(x,y))$.

Comparing \Eref{scaling_C} to \Eref{scaling_H} shows that both ensembles
display the same scaling behaviour. The lack of this proof might explain
the critical remarks about the conditional ensemble in
\cite{LuebeckHeger:2003}. The finite size scaling for the two
observables, however, differs: Setting $\delta p=0$ and choosing $\lambda$ so
that $a_L L \lambda^{-\nu_\perp}=1$ \Eref{scaling_C} gives
\begin{equation}
\ave{\rho^n}_C(0;L) = \frac{a_h a_n a_L^{d-\beta_n/\nu_\perp}}{a_V a_t}
L^{-\beta_n/\nu_\perp} R_n^C(0,1) \ .
\elabel{fss_C}
\end{equation}
Because $\ave{\rho^n}_C(0;L)$ is non-zero by definition but bound from
above, $R_n^C(0,1)$ must be finite, so that $\ave{\rho^n}_C(0;L)\propto
L^{-\beta_n/\nu_\perp}$ as one would expect from equilibrium critical
phenomena. Similarly, one finds
\begin{multline}
\ave{\rho^n}_H(0,h;L) = a_n a_L^{-\beta_n/\nu_\perp}
L^{-\beta_n/\nu_\perp} R_n^H(0, a_h h (a_L L)^{\sigma/\nu_\perp}, 1)\\
= a_n a_L^{-\beta_n/\nu_\perp}
L^{-\beta_n/\nu_\perp} a_h h (a_L L)^{\sigma/\nu_\perp}
\tilde{R}_n(0,1)
\elabel{fss_H}
\end{multline}
from \Eref{scaling_H} and \Eref{linear_Rn}. Again, the activity
$\ave{\rho^n}_H(0,h;L)$ is bound from above and does not vanish at finite
external field, so that $\tilde{R}_n(0,1)$ is finite, and therefore
$\ave{\rho^n}_H(0,h;L)\propto L^{(\sigma-\beta_n)/\nu_\perp}$. This
rather unusual finite size scaling behaviour of the activity is
documented in the literature \cite{Luebeck:2004} (in particular
Figures~33 and 47).
The effective finite size
scaling exponent for $n=1$ is $(\sigma-\beta)/\nu_\perp=\gamma/\nu_\perp$, which
is the finite size scaling exponent of the susceptibility. Indeed, for
small external fields at $\delta p=0$ and in finite systems, one expects
$\ave{\rho}_H=h \chi_H$.

\Eref{triviality} implies that, under the general conditions stated earlier,
\emph{all} moments vanish linearly in the external field. The Binder
cumulant
\begin{equation}
Q(\delta p, h, L) = 1 - \frac{\ave{\rho^4}_H}{3 \ave{\rho^2}_H^2}
\elabel{def_Q}
\end{equation}
therefore diverges like $h^{-1}$. 
In
\cite{LuebeckHeger:2003} it has been suggested that a divergent Binder
cumulant is a ``characteristic feature of all absorbing phase
transitions'', but in the light of the above derivation, it appears
rather like a generic feature of the particular ensemble. Moreover,
replacing $\ave{\rho^n}_H$ by the conditional averages
$\ave{\rho^n}_C$ in \Eref{def_Q} renders the Binder cumulant a
universal moment ratio again and reinstates it as a ``very useful
method'' to identify the transition.
The same applies to lower moment ratios, which can be determined with
higher accuracy, such as
$\ave{\rho}^2_C/\ave{\rho^2}_C$. 

It is now also
clear why the slightly more complicated moment ratio proposed in
\cite{LuebeckJanssen:2005} 
\begin{equation}
U=\frac{
\ave{\rho^2}_H \ave{\rho^3}_H 
 -
\ave{\rho}_H \ave{\rho^2}^2_H 
}
{
\ave{\rho}_H \ave{\rho^4}_H 
 -
\ave{\rho}_H \ave{\rho^2}^2_H 
}
\end{equation}
converges to a finite value: Both, denominator as
well as numerator are, to leading order, quadratic in $h$. In fact
\begin{equation}
\lim_{h\to0} U(\delta p, h, L) = 
\lim_{h\to0}
\frac{
\ave{\rho^2}_H \ave{\rho^3}_H
}
{
\ave{\rho}_H \ave{\rho^4}_H
}
=
\frac{
\ave{\rho^2}_C \ave{\rho^3}_C
}
{
\ave{\rho}_C \ave{\rho^4}_C
}
\end{equation}
where the first equality holds
because $\ave{\rho}_H \ave{\rho^2}^2_H\in\OC(h^3)$, and the second
because the remaining terms carry the same prefactor, namely $(h V(L)
\ave{t_m}_C(\delta p;L))^2$, see \Eref{triviality}.
The findings above are also confirmed by the exact results in
\cite{LuebeckJanssen:2005} (the RHS of Eq.~(21) in
\cite{LuebeckJanssen:2005} should read
$2/(\pi x)$ for $x\to0$).

The linearity of $\ave{\rho}_H$ in $h$, \Eref{linear_Rn}, contradicts
the mean field result $\ave{\rho}_H(\delta p=0, h)=\sqrt{h}$
\cite{Luebeck:2004}, which is, however, consistent with mean field
theory not allowing a proper distinction between conditional ensemble and
external field ensemble.

The limitations of \Eref{triviality} become clearer when comparing to the
standard scaling assumption $\lim_{L\to\infty} \ave{\rho}_H(\delta
p=0,h;L) \propto h^{\beta/\sigma}$ \cite{Hinrichsen:2000}.
\Eref{triviality} seems to suggest that $\beta/\sigma=1$ which clearly
is not the case. However, \Eref{triviality} applies only where the
waiting time
$(h V)^{-1}\propto L^{-d}$ is large compared to $\ave{t_m}_C\propto
L^{\sigma/\nu_\perp - d}$, which cannot be the case if
$\sigma/\nu_\perp>0$ and the thermodynamic limit taken at finite field
$h$. It is worth noting, however, that \Eref{linear_Rn} relies solely on the
Poissonian nature of the external field.

Finally, I want to discuss briefly two features of
conditional ensembles as they often appear in the literature, namely the
initial condition and the discounting of a transient. The latter was
introduced as $T$ in \Eref{tech_def_cond}, by defining the conditional
$n$th moment as an average taken from $t=T$ to $t\to\infty$. For any
non-vanishing transient $T>0$ the strict identity of conditional and
external field ensemble breaks down. To restore the above results,
more elaborate arguments are needed, based on the length of transient,
its contribution to the average and its scaling.

Similarly, there is no strict identity of the two ensembles,
if the conditional ensemble is based on initial conditions
which do not correspond to configurations reached when the external
field operates on inactive configurations in the external field
ensemble. 
On the other hand, one might hope that the introduction of a transient
erases the dependence on the initial condition and that asymptotically,
the effect of the transient and the initial condition becomes negligible. 

\section{Summary}
To summarise, I address the issue of the relation between
the two ensembles usually considered in absorbing state phase
transitions.  It turns out that the ensemble of ``activity conditional
to activity'' can be derived from the ensemble obtained by applying an
external field and vice versa, provided that the avalanche duration is
small compared to the waiting time between two avalanches.  Technically,
both methods can be implemented in (almost) the same way. Because moment
ratios, such as the Binder cumulant, remain finite in the conditional
ensemble and because moments of the activity in the presence of an
external field all vanish linearly in the external field, the
conditional ensemble might be more comfortable to study. 

\section*{Acknowledgements}
The author would like to thank Alastair Windus and Sven L{\"u}beck for
useful discussions and the RCUK for support.

\bibliography{articles,books}

\end{document}